\documentclass[prb,showpacs,twocolumn,preprintnumbers,amsmath,amssymb,floatfix]{revtex4}
\usepackage[dvips]{graphicx}
\usepackage[latin1]{inputenc}
\begin{document}
\draft

\hyphenation{a-long}

\title{Vortex dynamics and irreversibility line in optimally
doped SmFeAsO$_{0.8}$F$_{0.2}$\\ from ac susceptibility and
magnetization measurements}

\author{G. Prando,$^{1,2}$ P. Carretta,$^{1}$
R. De Renzi,$^{3}$ S. Sanna,$^{1}$ A. Palenzona,$^{4}$ M.
Putti,$^{4}$ M. Tropeano$^{4}$}

\address{$^{1}$Department of Physics ``A. Volta,'' University of Pavia-CNISM, I-27100 Pavia, Italy}
\address{$^{2}$Department of Physics ``E. Amaldi,'' University of Roma Tre-CNISM, I-00146 Roma, Italy}
\address{$^{3}$Department of Physics, University of Parma-CNISM, I-43124 Parma, Italy}
\address{$^{4}$CNR-INFM-LAMIA and University of Genova, I-16146 Genova, Italy}

\widetext

\begin{abstract}
Ac susceptibility and static magnetization measurements were
performed in the optimally doped SmFeAsO$_{0.8}$F$_{0.2}$
superconductor. The field - temperature phase diagram of the
superconducting state was drawn and, in particular, the features of
the flux line lattice derived. The dependence of the intra-grain
depinning energy on the magnetic field intensity was derived in the
thermally-activated flux creep framework, enlightening a typical
$1/H$ dependence in the high-field regime. The intra-grain critical
current density was extrapolated in the zero temperature and zero
magnetic field limit, showing a remarkably high value $J_{c0}(0)
\sim 2 \cdot 10^{7}$ A/cm$^{2}$, which demonstrates that this
material is rather interesting for the potential future
technological applications.
\end{abstract}


\maketitle

\narrowtext

\section{Introduction}

Several properties of iron-based 1111 oxy-pnictide superconductors
like, for instance, the high crystallographic and superconducting
anisotropy and large penetration depths make them similar to cuprate
high-$T_{c}$ materials. Since their discovery,\cite{Kam06,Kam08}
several studies have been performed in order to clarify their
intrinsic microscopic properties.\cite{Lum10,Joh10} The attention
has been mainly focussed on the bosonic coupling mechanism of the
superconducting electrons, on the features of the spin density wave
magnetic phase characterizing the parental and lightly
doped-compounds and its possible coexistence with
superconductivity,\cite{Lue09,San09,San10} and on the interaction
between localized magnetism from RE ions and itinerant electrons
onto FeAs bands.\cite{Pra10,Pou08,Sun10} A common hope is that a
full comprehension of these topics in oxy-pnictide superconductors
could also allow one to answer several open questions on the
cuprates.

At the same time, other analogies with cuprates possibly
characterize 1111 oxy-pnictides as interesting materials for
technological applications, like small coherence lengths (and,
correspondingly, high values of upper critical fields) besides the
high values of the superconducting critical temperature $T_{c}$. In
this respect, studies of macroscopic properties like critical fields
and critical depinning currents are of the utmost importance.
Namely, the investigation of the dynamical features of the flux line
lattice and of the so-called irreversibility line, typically
investigated by means of both resistance and ac susceptibility
measurements, is in order. Those measurements allow to further check
the validity of the theories used to model the mixed state of
cuprate materials and, in particular, the vortices motion and its
relationship with the possible pinning mechanisms.\cite{Bla94}
Several works reporting magnetoresistance,\cite{Mol10,Lee10,Sha10}
modulated microwave absorption\cite{Pan10} and dc
magnetization\cite{Jo09,Vdb10,Yam08,Ahm09} examining the flux line
lattice dynamics in 1111 oxy-pnictides have already been published
in the last two years. To our knowledge, no study of the flux line
lattice by means of ac susceptibility measurements has been
published yet.

This paper deals with the field, temperature and frequency
dependences of ac susceptibility in optimally doped
SmFeAsO$_{0.8}$F$_{0.2}$, which is one of the compounds showing the
highest $T_{c}$ among all the iron-based superconductors. Although
no large enough single-crystals are available and our data refer to
unoriented powder samples, the power of the ac susceptibility
technique allowed us to deduce several intrinsic features of the
mixed state of the superconductor. The magnetic field ($H$)
behaviour of the irreversibility line was obtained, allowing to draw
(together with dc magnetization data) a detailed phase diagram of
the flux line lattice. The $H$ dependence of the intra-grain
effective depinning energy ($U_{0}$) was investigated, evidencing
the characteristic crossover from a single-vortex-dynamics to a
collective-dynamics ($U_{0} \propto 1/H$) at a field $H \sim 0.5$ T.
An estimate of the intra-grain critical depinning current density in
the limit of vanishing temperature and magnetic field was also
deduced, giving the remarkably high value of $J_{c0}(0) \sim 2 \cdot
10^{7}$ A/cm$^{2}$. This result is of great importance in
characterizing SmFeAsO$_{0.8}$F$_{0.2}$ as a superconductor suitable
for technological applications.

\section{Experimentals and main results}\label{SectExp}

SmFeAsO$_{0.8}$F$_{0.2}$ was prepared by solid state reaction at
ambient pressure from Sm, As, Fe, Fe$_{2}$O$_{3}$ and FeF$_{2}$.
SmAs was first synthesized from pure elements in an evacuated,
sealed glass tube at a maximum temperature of 550°C. The final
sample was synthesized by mixing SmAs, Fe, Fe$_{2}$O$_{3}$ and
FeF$_{2}$ powders in stoichiometric proportions, using uniaxial
pressing to make powders into a pellet and then heat treating the
pellet in an evacuated, sealed quartz tube at 1000°C for 24 hours,
followed by furnace cooling. The sample was analyzed by powder X-ray
diffraction in a Guinier camera, with Si as internal standard. The
powder pattern showed the sample to be single phase with two weak
extra lines at low angle of the SmOF extra phase. The lattice
parameters were $a = 3.930(1)$ Å and $c = 8.468(2)$ Å.

\begin{figure}[htbp]
\vspace{6.6cm} \includegraphics{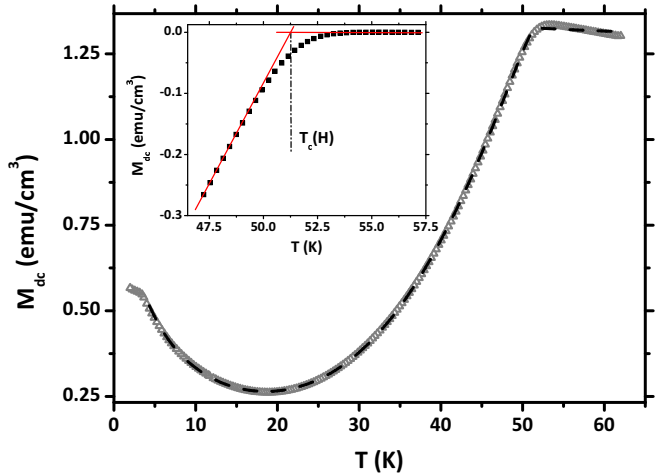} \caption{\label{FigDC1v5T}(Color online)
Temperature dependence of static magnetization $M_{dc}$ upon
field-cooling the sample at $H = 1.5$ T. Dashed lines are the best
fitting functions according to Eq. \ref{EqScTrans} (see later on in
the text). Inset: estimate of $T_{c}$($H = 1.5$ T) after the
subtraction of the paramagnetic contribution.}
\end{figure}

Static magnetization $M_{dc}$ measurements were performed by means
of a Quantum Design MPMS-XL7 SQUID magnetometer. The temperature $T$
dependence of $M_{dc}$ upon field-cooling (FC) the sample was
monitored at different applied magnetic fields up to 7 T.
Representative raw $M_{dc}$ vs. $T$ curves are shown in Fig.
\ref{FigDC1v5T}.
The superconducting (SC) response, with onset around $T_{c} \simeq
52$ K, is found to be superimposed to a paramagnetic contribution
associated with Sm$^{3+}$ ions. Clear kinks in the magnetization
curves can be observed at $T_{N} \simeq 4$ K, evidencing the
antiferromagnetic transition of the Sm$^{3+}$ magnetic
moments.\cite{Rya09} The field dependence of the SC transition
temperature $T_{c}(H)$ was deduced by first subtracting the linear
extrapolation of the Sm$^{3+}$ paramagnetic contribution in a few-K
region around the SC onset from the raw data. The transition
temperatures were then estimated from the intersection of two linear
fits of the resulting curves above and below the onset (see Fig.
\ref{FigDC1v5T}, inset). $T_{c}(H)$ behaviour was also deduced by
means of magnetoresistance measurements upon the application of
magnetic fields up to 9 T, showing a behaviour analogous to what
observed in SmFeAsO$_{1-x}$F$_{x}$ compounds from the same bath but
with lower $x$ concentrations of F$^{-}$.\cite{Pal09}

The onset of the diamagnetic contribution and its dependence on the
applied external magnetic field was also investigated by means of a
Quantum Design MPMS-XL5 SQUID ac susceptometer. Measurements were
performed with an alternating field in the range $H_{ac} =$
($0.0675$ -- $1.5$) $\cdot 10^{-4}$ T parallel to the static field
$H$, which varied up to 5 T. The ac field frequency ranged from $37$
to $1488$ Hz. The diamagnetic onset temperature was estimated from
$\chi^{\prime}$ vs. $T$ curves by means of the same procedure shown
in the inset of Fig. \ref{FigDC1v5T}.
\begin{figure}[htbp]
\vspace{6.6cm} \includegraphics{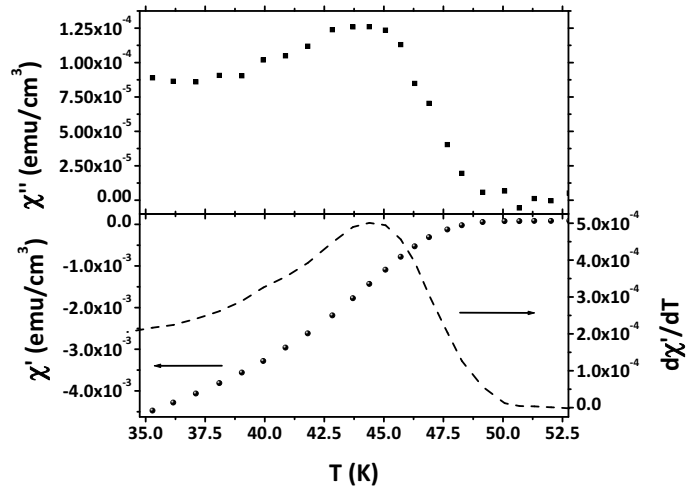} \caption{\label{FigAC}Temperature
dependence of the real and imaginary components of ac susceptibility
$\chi^{\prime}$ and $\chi^{\prime\prime}$ (bottom and top panels,
respectively). The alternating field amplitude $H_{ac} = 1.5 \cdot
10^{-4}$ T at the frequency $\nu_{m} = 37$ Hz is superimposed to a
larger static field $H = 1.5$ T. The dashed line in the bottom panel
represents the first derivative of $\chi^{\prime}$ with respect to
temperature.}
\end{figure}

An accurate examination of ac susceptibility data as a function of
$\nu_{m}$ allowed us to obtain further information on the dynamical
properties of the flux line lattice (FLL). It is well known, in
fact, that a peak in $\chi^{\prime\prime}$ vs. $T$ curves associated
with a maximum in the energy dissipation inside the sample appears
at a temperature $T_{p}$ slightly lower than the diamagnetic onset
temperature in $\chi^{\prime}$. Correspondingly, at the same
temperature $T_{p}$ the $\chi^{\prime}$ vs. $T$ curve displays a
maximum in its first derivative\cite{Kes89} (see exemplifying raw
data in Fig. \ref{FigAC}, lower panel). Several works in the past
decades have tried to clarify the origin of the
$\chi^{\prime\prime}$ peak. One of the possible interpretations
relies on the Bean's critical state model\cite{Bea62,Bea64} and
associate the peak in $\chi^{\prime\prime}$ with the flux front
reaching the centre of the sample. In this case, the peak
temperature $T_{p}$ should not depend on the measurement frequency
$\nu_{m}$ and a strong dependence on sample dimensions and ac field
amplitude $H_{ac}$ are predicted.\cite{Fri95} Another interpretation
relies on the modification of the skin depth, due to the
superconductor resistivity in the thermally-assisted flux flow
(TAFF) regime, with respect to the London penetration depth. In this
case $T_{p}$ should strongly depend on the measurement frequency
$\nu_{m}$ while no dependence on the ac field amplitude $H_{ac}$ is
predicted.\cite{Kes89,Fri95,Cle91,Ges91,Bra93,Gom97} Considering the
frequency dependence of the $\chi^{\prime\prime}$ peak, another
interesting interpretation has been associated with a resonant
absorption of energy obtained when the inverse of $\nu_{m}$ matches
the characteristic relaxation time $\tau_{c}$ of the FLL at
$T_{p}$,\cite{Pal90,Tin91} namely
\begin{equation}
    2 \pi \nu_{m} \tau_{c|_{T = T_{p}}} = 1.
\end{equation}
In this case, the underlying theory is the more general framework of
the thermally activated creep of flux lines between different
metastable minima of pinning potentials.\cite{And64,Tin96}

At temperatures lower than $T_{p}$, other broader contributions to
both $\chi^{\prime}$ and $\chi^{\prime\prime}$ were found and
interpreted as arising from granularity of the powder sample and, in
particular, to intergranular Josephson weak links between different
grains.\cite{Gom97,Nik89,Nik95} In cuprate materials, from the
analysis of the low-temperature peak in $\chi^{\prime\prime}$ and,
in particular, of its frequency dependence, the depinning energy
barrier associated with grain boundaries was
extracted.\cite{Nik89,Kum95,Mul90} Strong granularity has been
observed also in iron-based pnictide materials. On samples prepared
with the same procedure a small (though not negligible)
intergranular critical density current has been evaluated by a
remanent magnetization analysis.\cite{Yam11} However, it has been
determined that the main contribution to the magnetization curve
comes from intagranular currents. By considering our data just in
the $T$ region close enough to the diamagnetic onset, then, we will
be focussing only on the intra-grain intrinsic dynamical properties.
The stronger signal amplitude, moreover, made it reasonable to
investigate the peak in $\chi^{\prime}$ derivative instead of the
maximum in $\chi^{\prime\prime}$.

\begin{figure}[htbp]
\vspace{6.6cm} \includegraphics{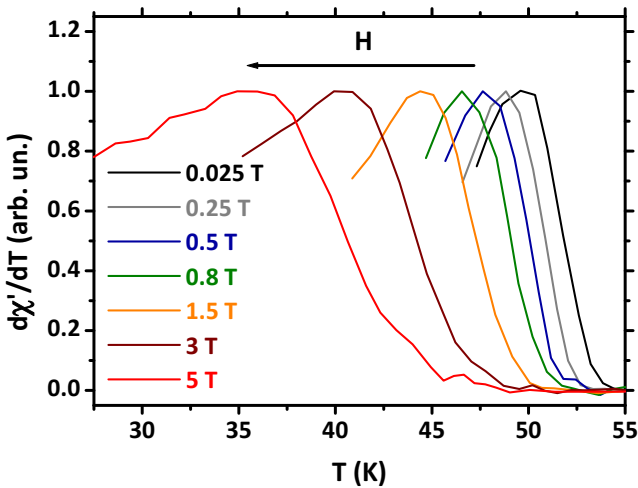} \caption{\label{FigACder}(Color online)
Temperature dependence of the first derivative of the real component
of ac susceptibility $\chi^{\prime}$ ($H_{ac} = 1.5 \cdot 10^{-4}$
T, $\nu_{m} = 37$ Hz) with respect to temperature at different
applied magnetic fields. All the single curves were independently
normalized to the corresponding maximum derivative value.}
\end{figure}
Fig. \ref{FigACder} shows the temperature evolution of the
normalized first derivative of the real component $\chi^{\prime}$ of
the ac susceptibility upon the application of different values of
static magnetic field $H$. In these measurements, both the
alternating field and the working frequency were kept fixed at
$H_{ac} = 1.5 \cdot 10^{-4}$ T and $\nu_{m} = 37$ Hz respectively.
The effect of increasing $H$ is a clear shift of $T_{p}$ towards
lower values. At each applied $H$ a clear dependence of $T_{p}$ on
$\nu_{m}$ was evidenced, as discussed later on. Some scans with
$H_{ac}$ values in the range ($0.0675$ -- $1.5) \cdot 10^{-4}$ T
were also performed at the representative values $H = 0.025, 0.25$
and $5$ T (data not shown). Within the experimental error, no
dependence of $T_{p}$ on $H_{ac}$ was evidenced.

\section{Analysis and discussion}

The raw $M_{dc}$ vs. $T$ data reported in Fig. \ref{FigDC1v5T} were
fitted by the function (see dashed curves in Fig. \ref{FigDC1v5T})
\begin{eqnarray}\label{EqScTrans}
    M_{dc}(T,H) &=& M_{sc} \left[1 -
    \left(\frac{T}{T_{c}}\right)^{\alpha}\right]^{\beta}
    +\nonumber\\ &&
    + C_{cw}\frac{H}{T - T_{N}} + M_{0}(H)
\end{eqnarray}
where the first term is the diamagnetic Meissner response
(empirically represented by a two-exponents mean-field function) and
the second one is the Curie-Weiss paramagnetic contribution. The
last term accounts for all the sources of $T$-independent magnetism,
ranging from Pauli- and Van-Vleck-like paramagnetism to a small
contribution of magnetic impurities (e.g. Fe$_{2}$As).\cite{Cim09} A
detailed analysis of the results of the fitting procedure according
to Eq. \ref{EqScTrans} will be presented in another
work.\cite{Pra11}

\begin{figure}[htbp]
\vspace{6.6cm} \includegraphics{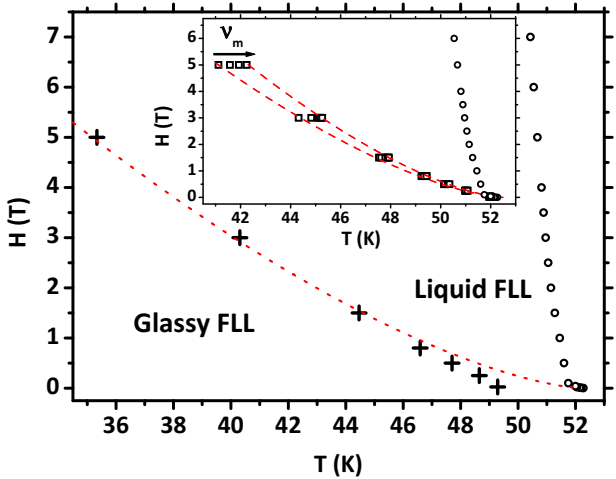} \caption{\label{FigPhDiaSubm}(Color
online) Phase diagram associated with the flux line lattice. A
glassy phase of FLL is separated from a liquid phase by the
irreversibility line obtained from the field dependence of the
maximum slope of $\chi^{\prime}$ (plus signs (+), relative to
$H_{ac} = 1.5 \cdot 10^{-4}$ T, $\nu_{m} = 37$ Hz). The red dotted
line is a best fit according to Eq. \ref{EqPowLaw}. Circles
($\circ$) track the behaviour of $H_{c2}$ as measured from dc
magnetization data. Inset: comparison of the diamagnetic onset
measured from dc magnetization ($\circ$) and ac susceptibility
($\Box$) with $H_{ac} = 1.5 \cdot 10^{-4}$ T and $\nu_{m} = 37$ --
$1488$ Hz. The red dashed lines are best fits of data at $\nu_{m} =
37$ Hz and $\nu_{m} = 1488$ Hz according to Eq. \ref{EqPowLaw}.}
\end{figure}

Results from both SQUID magnetometry and ac susceptibility are
summarized in the phase diagram shown in Fig. \ref{FigPhDiaSubm}.

From the $T_{c}(H)$ data obtained from $M_{dc}$ vs. $T$ curves (see
Fig. \ref{FigDC1v5T}) it was possible to derive the temperature
dependence of the upper critical field $H_{c2}$ (see open circles in
Fig. \ref{FigPhDiaSubm}). A linear fit of the $H_{c2}$ vs. $T$ data
deduced from magnetization data gives a slope $d H_{c2}/d T = 7.47
\pm 0.15$ T/K, in agreement with what was found in compounds of the
same family from magnetoresistivity data,\cite{Pal09,Jar08} even if
much lower slope values were reported from calorimetric measurements
on single crystals of Nd-based 1111 superconductors.\cite{Wel08}
Then, in the simplified assumption of single-band superconductivity,
through the Werthamer, Helfand and Hohenberg relation\cite{Wer66}
\begin{equation}\label{EqWHH}
    H_{c2}(T = 0 K) \simeq 0,693 \cdot T_{c} \left|\frac{d H_{c2}}{d
    T}\right|_{T \simeq T_{c}}
\end{equation}
it is possible to estimate $H_{c2}$ $\simeq 270$ T in the limit of
vanishing $T$.

A comparison between the diamagnetic onset temperature as obtained
from ac susceptibility at different frequencies (see exemplifying
raw data in Fig. \ref{FigAC}) and the one obtained from static
magnetization is plotted in the inset of Fig. \ref{FigPhDiaSubm}.
The onset in ac data was systematically found at lower temperatures
than the corresponding dc diamagnetic onset. Dashed lines represent
the empirical power-law fitting function
\begin{equation}\label{EqPowLaw}
    H \propto (1 - T/T_{c})^{\beta}, \; \beta = 3/2
\end{equation}
well describing experimental data at each value of the ac field
frequency $\nu_{m}$. Such a functional form characterized by $\beta
= \frac{4}{3}$ -- $\frac{3}{2}$ is known to describe the
irreversibility line in the $H-T$ phase diagram of
cuprates.\cite{Mal88}

Plus signs in the phase diagram in Fig. \ref{FigPhDiaSubm} refer to
the points $T = T_{p}$ of maximum slope of $\chi^{\prime}$ vs. $T$
curves ($H_{ac} = 1.5 \cdot 10^{-4}$ T and $\nu_{m} = 37$ Hz. See
Fig. \ref{FigACder}) corresponding, as already explained in sect.
\ref{SectExp}, to the maximum of $\chi^{\prime\prime}$ vs. $T$
associated with intrinsic intra-grain losses. These data divide the
phase diagram into two main regions, following the interpretation of
the $\chi^{\prime\prime}$ peak in term of resonant absorption of
energy in a thermally activated flux creep model.\cite{Pal90} In the
high-$T$ and high-$H$ region the flux lines are in a reversible
state, that is, they are responding to the external ac perturbation
(liquid FLL). On the other hand, in the low-$T$ and low-$H$ region
vortices are arranged in a glassy-like frozen FLL that gives rise to
a non-reversible response and to dissipation, linked to the non-zero
values assumed by $\chi^{\prime\prime}$. $T_{p}$ vs. $H$ points
associated with the lowest accessible frequency $\nu_{m}$ are thus
expected to belong to the irreversibility line (or de Almeida -
Thouless line) of the FLL phase diagram. As in the case of the
diamagnetic onset in $\chi^{\prime}$ (see Fig. \ref{FigPhDiaSubm},
inset), Eq. \ref{EqPowLaw} nicely fits the field dependence of
$T_{p}$ in the $H > 0.8$ T limit (see the dotted line in Fig.
\ref{FigPhDiaSubm}).

\begin{figure}[htbp]
\vspace{6.6cm} \includegraphics{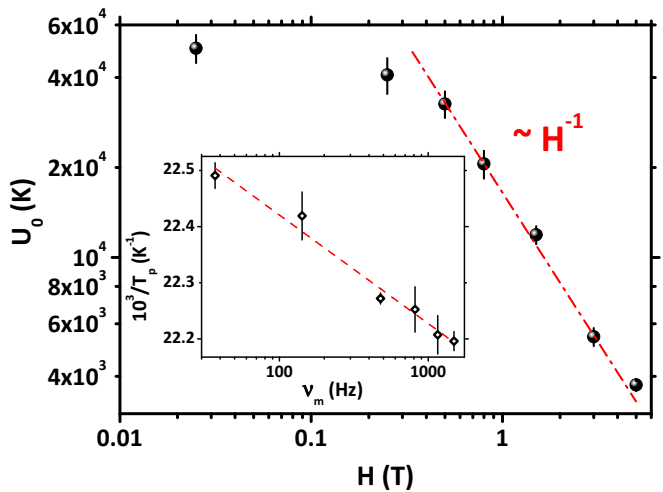} \caption{\label{FigDepBarr}(Color online)
Magnetic field dependence of the effective depinning energy barrier.
The red dashed-dotted line is a best fit of data according to a
$1/H$ dependence for $H \gtrsim 0.5$ T. Inset: frequency dependence
of $1/T_{p}$. Data refer to $H = 1.5$ T with the alternating field
amplitude fixed to $H_{ac} = 1.5 \cdot 10^{-4}$ T. The red dashed
line is a best fit of the experimental data according to Eq.
\ref{EqLogDep}.}
\end{figure}

A logarithmic dependence of $1/T_{p}$ vs. $\nu_{m}$ at every fixed
$H$ is evidenced over the explored frequency range (see, for
example, $H = 1.5$ T data in the inset of Fig. \ref{FigDepBarr}).
Data can then be fitted within a thermally-activated framework by
the formula (red dashed line in the inset of Fig. \ref{FigDepBarr})
\begin{equation}\label{EqLogDep}
    \frac{\nu_{m}}{\nu_{0}} = \exp\left(-\frac{U_{0}(H)}{k_{B}
    T_{p|_{\nu = \nu_{m}}}}\right)
\end{equation}
from which it can be observed that the logarithmic behaviour of
$1/T_{p}$ is mainly controlled by the parameter $U_{0}$, playing the
role of an effective depinning energy barrier in a
thermally-activated flux creep model. The parameter $\nu_{0}$ takes
the meaning of a intra-valley characteristic frequency associated
with the motion of the vortices around their equilibrium position in
the pinning centers.

The advantage of extracting the value of $U_{0}$ from ac
susceptibility data, if compared for instance to magnetoresistance
data, is that the former is an almost isothermal estimate.
Temperature, in fact, varies at most by $1$ K degree as a function
of $\nu_{m}$ at the highest applied $H$ (see Fig. \ref{FigDepBarr},
inset) so that it is possible to determine $U_{0}$ at a temperature
$T^{*}(H)$ with a maximum uncertainty of $0.5$ K. This fact will be
of interest when deriving the critical current density value, as it
will be shown shown later on.

Data at different static magnetic fields can be fitted according to
this model giving the results reported in Fig. \ref{FigDepBarr},
where the $H$ dependence of the effective depinning energy barrier
$U_{0}$ is shown. Beyond an overall sizeable reduction of $U_{0}$
with increasing $H$, a crossover between two different regimes can
be clearly observed at $H \gtrsim 0.5$ T. At high fields the data
are well described by a $1/H$ dependence, a well-known result in
high-$T_{c}$ cuprate superconductors, observed by means of several
techniques, ranging from NMR\cite{Rig98} to ac
magnetometry\cite{Emm91} and magnetoresistivity\cite{Pal90}. A naive
explanation of this behaviour can be obtained in terms of the
balance between the Gibbs free energy of the system and the energy
required for the motion of a flux lines bundle.\cite{Tin88} In this
framework, the crossover between the two different trends of $U_{0}$
vs. $H$ shown in Fig. \ref{FigDepBarr} can be interpreted as the
transition from a basically single-flux line response at low $H$
values to a collective response of vortices for $H > 0.5$ T. A
similar phenomenology was recently reported from magnetoresistivity
data on a single crystal of O-deficient SmFeAsO$_{0.85}$ and on
powder samples of La-based and Ce-based 1111
superconductors.\cite{Lee10,Sha10} The crossover between different
regimes was observed at $H \simeq 1$ T in La-based samples and at
much higher magnetic fields in Ce-based samples and in
SmFeAsO$_{0.85}$ ($H \simeq 3$ T). However, $U_{0}$ values are
typically 1 order-of-magnitude lower in La- and Ce- based
superconductors. Comparing the sets of data for $U_{0}$ derived from
magnetoresistivity and here from ac susceptibility in Sm-based
samples, the numerical agreement is very good for $H \gtrsim 1$ T.

\begin{figure}[htbp]
\vspace{6.6cm} \includegraphics{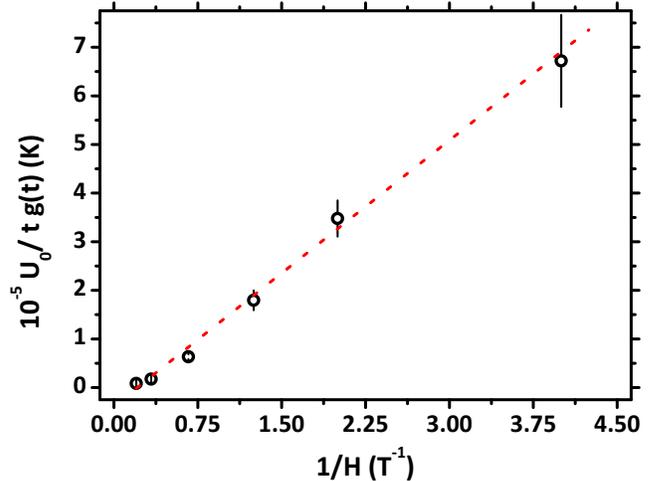} \caption{\label{FigJc}(Color online) Plot
of the effective energy barrier normalized with respect to the
reduced temperature function $t g(t)$ (see text) as a function of
the inverse applied static field. The slope of the linear fit (see
red dotted curve) is directly proportional to the critical current
density extrapolated at $T = 0$ K and $H = 0$ T.}
\end{figure}

In the $1/H$ regime, Tinkham\cite{Tin88} extended previous works by
Yeshurun and Malozemoff\cite{Mal88,Yes88} showing that the relation
for the normalized effective depinning energy barrier
\begin{equation}\label{EqDep}
    \frac{U_{0}(t,H)}{t} \simeq \frac{K J_{c0}(0)}{H} g(t)
\end{equation}
holds also for granular samples. Here $U_{0}$ is expressed in K
degrees, $t$ is the reduced temperature $t \equiv T/T_{c}$, $g(t)
\equiv 4 \left(1-t\right)^{3/2}$, $J_{c0}(t)$ quantifies the
critical current density at $H = 0$ and $T = t T_{c}$, while the
constant $K$ is defined as $K \equiv 3 \sqrt{3} \Phi_{0}^{2} \beta/2
c$, $\Phi_{0}$ being the flux quantum, $c$ the speed of light and
$\beta$ a numerical constant close to unit value. Eq. \ref{EqDep} is
derived in the simplified assumption of a two-fluid
model.\cite{Tin88} By assuming that this empirical scenario can
describe also the system under current investigation, from ac
susceptibility it is possible to directly extrapolate the value of
$J_{c0}(0)$. $U_{0}(H)$, as already noticed above, is almost
isothermally estimated and can then be expressed as
$U_{0}(t^{*},H)$. By now plotting $U_{0}/t^{*} g(t^{*})$ as a
function of $1/H$ (see Fig. \ref{FigJc}), from a linear fit of data
it is possible to extract from Eq. \ref{EqDep} the value $J_{c0}(0)
= \left(2.25 \pm 0.05\right) \cdot 10^{7}$ A/cm$^{2}$, having
assumed $\beta = 1$. This rather high value is in agreement with
estimates of the critical current density $J_{c}$ evaluated by
inductive measurements in similar Sm-based samples\cite{Yam11} and
also with the direct measurement of this quantity in a Sm-based 1111
single crystals in the $T \rightarrow 0$ K and $H \rightarrow 0$ T
limit.\cite{Mol10}

\section{Conclusions}

The $H$ -- $T$ phase diagram of the flux line lattice in a powder
sample of optimally-doped SmFeAsO$_{0.8}$F$_{0.2}$ was investigated
by means of both ac and dc susceptibility measurements. The
irreversibility line separating a liquid from a glassy phase was
deduced and the activation depinning energy $U_{0}$ as a function of
the external magnetic field derived in the framework of a
thermally-activated flux creep theory. A $1/H$ dependence of $U_{0}$
for $H \gtrsim 0.5$ T, typical of collective motion of flux lines,
has been evidenced. From the $U_{0}$ vs. $H$ behaviour a value of
$J_{c0}(0) \sim 2 \cdot 10^{7}$ A/cm$^{2}$ has been extrapolated for
the critical depinning current at both zero field and zero
temperature. From this result we confirm that high intrinsic
critical depinning current density values seem to be a peculiar
feature of these systems, possibly making them good candidates for
technological applications.

\begin{acknowledgements}
M. J. Graf, A. Rigamonti and L. Romanò are gratefully acknowledged
for stimulating discussions. We thank C. Pernechele and D. Zola for
useful help and suggestions about ac susceptibility measurements.
\end{acknowledgements}



\end{document}